\documentstyle[preprint,epsf]{jpsj}

\def\runtitle{Correlation Effects on Optical Conductivity of FeSi}
\def\runauthor{Kentaro {\sc Urasaki} and Tetsuro {\sc Saso}}

\title
{
Correlation Effects on Optical Conductivity of FeSi
}

\author
{
Kentaro {\sc Urasaki}\footnote{e-mail address: 
kentaro@krishna.th.phy.saitama-u.ac.jp}
and Tetsuro {\sc Saso}\footnote{e-mail address: 
saso@phy.saitama-u.ac.jp}
}

\inst
{
Department of Physics, Saitama University, Urawa, 338-8570
}

\recdate
{
\today
}

\abst
{
Effects of electron correlation in FeSi are studied in terms of the two-band
Hubbard model with the density of states obtained from the band calculation.
Using the self-consistent second-order perturbation theory combined with the
local approximation, the correlation effects are investigated on the density of
states and the optical conductivity spectrum, which are found to reproduce the
experiments done by Damascelli {\it et al.} semiquantitatively. 
It is also found that the peak at the 
gap edge shifts to lower energy region by correlation effects, as is seen in
the experiments.
}

\kword
{
FeSi, optical conductivity, Kondo insulator, two-band Hubbard model
}

\begin{document}
\sloppy
\maketitle

FeSi is a well-known material for its anomalous properties that cannot be 
understood from conventional band theories. 
At low temperatures FeSi is a nonmagnetic semiconductor with 
a gap size of about 50 meV, whereas it has a magnetic moment 
and shows metallic behaviors at 
room temperatures.\cite{Jaccarino67,Watanabe63}
To explain the magnetic properties, Takahashi and Moriya\cite{Takahashi79} 
applied the spin fluctuation theory leading to an idea of thermally induced local moment,
 which was confirmed later by 
the neutron scattering experiment.\cite{Shirane87} 
Recently interest has been renewed by the suggestion\cite{Aeppli92} 
that FeSi can be viewed as a strongly correlated insulator 
or a Kondo insulator. 
Kondo insulators have been found in the $f$-electron systems 
and typical examples are YbB$_{12}$,\cite{Kasaya85} 
Ce$_3$Bi$_4$Pt$_3$\cite{Hundley90} and so on.
They have correlated $f$-bands and 
become insulators with small energy gaps at low temperatures.   
The correlation in FeSi may not be so strong as in these materials.
It will be, however, of interest to reinvestigate the properties of FeSi 
from a new point of view. 

Low temperature properties of FeSi 
have been studied intensively from this aspect. 
Especially, Schlesinger {\it et al.}\cite{Schlesinger93} 
measured the optical conductivity 
and reported that the gap ($\sim$50 meV) at low temperatures 
disappears at a temperature ($\sim$250 K) lower than the gap size, 
which they attributed to the correlation effect. 
The optical conductivity of FeSi was reported by several 
groups.\cite{Schlesinger93,Ohta94,Damascelli97,Paschen97} 
The gap opening of about 50 meV and the rapid temperature variation 
were commonly observed. 
Those optical measurements also provided us with a direct information about 
the dynamical structure of this system at low temperatures such as a precise gap size. 
Moreover, high-resolution photoemission measurement was done by
Saitoh {\it et al.},\cite{Saitoh95} who 
estimated the self-energy of the $d$-electrons 
and demonstrated a possible existence of the 
strong renormalization of the electronic states at low energies.
A recent experiment reported that a sign of the correlation effects is found 
in the tunneling spectroscopy measurements\cite{Fath98}, where the gap in the 
density of states (DOS) seems 
to be strongly temperature dependent. 
It was also reported that the low temperature transport measurements 
for Fe$_{1-x}$Co$_x$Si\cite{Chernikov97} and FeSi$_{1-x}$Al$_{x}$\cite{DiTusa98} 
suggest strong mass enhancement of carriers. 

In these contexts several theoretical attempts have been made 
to take the correlation effects into account. 
Among these attempts Fu and Doniach\cite{Fu95} 
proposed a two-band Hubbard model based on their band calculation.\cite{Fu94} 
Their calculation revealed the importance of the correlation effects on 
physical quantities but there 
seems to be some errors about the treatment of the self-energies. 
Using this model, we\cite{Urasaki98} also calculated physical 
quantities correctly, 
and showed that the temperature variations of 
the magnetic susceptibility and the optical conductivity 
are consistent with the experiments. 
From a viewpoint of the strongly correlated systems, 
Riseborough\cite{Riseborough98} calculated the angle-resolved-photoemission spectrum
of the Anderson lattice model in terms of the $1/N$ expansion. 

On the other hand several band calculations have been done for 
FeSi.\cite{Mattheiss93,Fu94,Jarlborg95,Galakhov95,Kulatov97} 
Each of them succeeded in giving the insulating ground state 
and showed that the minimum gap is given by the indirect one, 
but it is close to a direct one. 
It has been also demonstrated that 
the states near the Fermi level mainly consist of the 
$d$-electrons. 
However, somewhat different gap sizes (0.03-0.2 eV) were reported. 
There are some attempts to explain the physical properties of FeSi 
by extending the band calculations, 
e.g., Jarlborg\cite{Jarlborg99} investigated the thermal disorder effect 
and Anisimov {\it et al.}\cite{Anisimov96} applied the LDA+$U$ scheme. 
Fu {\it et al.}\cite{Fu94} calculated the optical conductivity 
without the correlation effect using their band calculation. 
As a result, the necessity of the correlation effects 
was underlined in order to reproduce the 
observed temperature variation. 
Ohta {\it et al.}\cite{Ohta94} also calculated the optical conductivity 
in the joint-DOS form from their band calculation and showed that the gap is 
filled with increasing temperature. 
In their calculation, the Drude contribution is estimated from the 
experimental DC resistivity and is added to the joint-DOS term separately. 
However, 
the flat part of the optical conductivity spectrum within the gap is not reproduced. 

Recently, Yamada {\it et al.}\cite{Yamada99} calculated the DOS within 
the LMTO-ASA method carefully, 
and obtained the gap size close to the experimental one.
Therefore, in the present Letter, 
we study the two-band Hubbard model using 
the Fe part of the DOS obtained from this 
band calculation and calculate the temperature variation of 
the optical conductivity 
to investigate the correlation effects. 
Moreover, we compare the calculated spectrum of the optical conductivity 
with the experimental data and estimate the gap size (indirect one) in the DOS 
of the real system. 

We start from the following two-band Hubbard Hamiltonian: 
\begin{eqnarray}
H=
& &\sum_{ij\sigma} 
(t^1_{ij}c_{i1\sigma}^\dagger c_{j1\sigma} +t^2_{ij}c_{2i\sigma}^\dagger 
c_{2j\sigma})\cr 
&+&U\sum_{i}(n_{i1\uparrow}n_{i1\downarrow}+ 
n_{i2\uparrow}n_{i2\downarrow})\cr 
&+&U_2\sum_{i}(n_{i1\uparrow}n_{i2\downarrow}+ 
n_{i2\uparrow}n_{i1\downarrow})\cr 
&+&U_3\sum_{i}(n_{i1\uparrow}n_{i2\uparrow}+ 
n_{i2\downarrow}n_{i1\downarrow})\cr 
&-&J\sum_{i}(c_{i1\uparrow}^\dagger c_{i1\downarrow} 
c_{i2\downarrow}^\dagger c_{i2\uparrow} + 
c_{i2\uparrow}^\dagger c_{i2\downarrow} 
c_{i1\downarrow}^\dagger c_{i1\uparrow} ), 
\end{eqnarray} 
where the $c^\dagger_{ia\sigma}(c_{ia\sigma})$ creates (destroys) 
an electron on site $i$ in band $a=1,2$ with spin $\sigma$. 
If the two bands are degenerate, this Hamiltonian is rotationally invariant in spin and real spaces when one chooses $U_2=U-J$ and $U_3=U-2J$,\cite{Parmenter73} where $U$ and $J$ denotes the Coulomb repulsion and the exchange interaction, respectively. We adopt this relation so as to reduce the number of parameters. 

The initial DOS obtained from the band calculation\cite{Yamada99} 
is displayed in Fig. 1 (see the solid line for $T=0$), 
where we widen the energy gap by 16 $\%$ so as to 
reproduce the gap edge of the optical conductivity at 4 K of the experiment. 
There is an ambiguity to determine the gap size $E_g$. 
If we extrapolate the steepest part of the DOS at 
the both sides of the gap neglecting the tails,
we obtain $E_g=75$ meV, 
whereas if we regard the gap as the region inside the tails of the gap edge, 
we obtain 60 meV. 
The Fermi level ($E_F=0$) is placed at the center of the gap at $T=0$, and 
we call the upper and the lower part of the DOS as the band 1 and 2, respectively. 
We introduce a cut off for each band so as to include one state per spin in each band. 
Then the band width for the band 1 is about 0.56 eV and about 0.63 eV for the band 2. 
Moreover, although the DOS is asymmetric, 
we assume that the chemical potential is temperature independent. 
This means that the electron number of this system varies with temperature. 
This assumption seems to be reasonable in the real system, 
where electrons and holes are supplied from other bands. 

Because the present interests are on the correlation effects 
in the low energy and low temperature region of this model, 
we adopt the self-consistent second-order perturbation theory (SCSOPT) 
combined with the local approximation.\cite{MullerHartmann89}
In SCSOPT, the self-energies $\Sigma_1^{(2)\sigma}(\omega)$ and 
$\Sigma_2^{(2)\sigma}(\omega)$ are calculated as
\begin{eqnarray}\label{eq:sigma}
\Sigma_1^{(2)\sigma}(\omega)&=&
\int\!\!\!\int\!\!\!\int^\infty_{-\infty}
d\varepsilon_1 d\varepsilon_2 d\varepsilon_3 \cr\cr
&&[U^2\rho_1^{-\sigma}(\varepsilon_1)
\rho_1^{\sigma}(\varepsilon_2)
\rho_1^{-\sigma}(\varepsilon_3) \cr\cr
&&+U_2^2\rho_2^{-\sigma}(\varepsilon_1)
\rho_1^{\sigma}(\varepsilon_2)
\rho_2^{-\sigma}(\varepsilon_3) \cr\cr
&&+U_3^2\rho_2^{\sigma}(\varepsilon_1)
\rho_1^{\sigma}(\varepsilon_2)
\rho_2^{\sigma}(\varepsilon_3) \cr\cr
&&+J^2\rho_2^{-\sigma}(\varepsilon_1)
\rho_2^{\sigma}(\varepsilon_2)
\rho_1^{-\sigma}(\varepsilon_3) 
]\cr\cr
\times&&
\hspace{-5mm}\frac{f(-\varepsilon_1)f(\varepsilon_2)f(\varepsilon_3)
+f(\varepsilon_1)f(-\varepsilon_2)f(-\varepsilon_3)}
{\omega+\varepsilon_1-\varepsilon_2-\varepsilon_3+{\rm i}\delta},\cr
\Sigma_2^{(2)\sigma}(\omega)&=&(1\leftrightarrow 2),
\end{eqnarray}
where
\begin{equation}\label{eq:rho}
\rho_a^\sigma(\omega)=
-\frac{1}{\pi}{\rm Im}G_a^{\sigma}(\omega+{\rm i}\delta)
\end{equation}
and
\begin{eqnarray}\label{eq:g}
G_a^\sigma(\omega)
&&=\frac{1}{N}\sum_{\bf k}G_a^{\sigma}({\bf k},\omega)\cr
&&=\int^{\infty}_{-\infty}d\varepsilon\rho^{0\sigma}_a(\varepsilon)
\frac{1}{\omega-\varepsilon-\Sigma^{(2)\sigma}_a(\omega)}.
\end{eqnarray}
Here, $N$ is the number of sites, $f(\varepsilon)$ the Fermi function 
and $\rho^{0\sigma}_a(\varepsilon)$ 
the DOS of band $a$ for the non-interacting case. 
We solve the set of equations (\ref{eq:sigma})-(\ref{eq:g}) self-consistently. 
To make numerical calculation easy, 
we take $\delta$ finite ($\delta=10^{-4}$) in eq. (\ref{eq:rho}) and 
convert these equations with the transformations\cite{MullerHartmann89}
\begin{eqnarray}
A_a^\sigma(\tau)&=&\int^\infty_{-\infty}d\epsilon
e^{-{\rm i}\tau\varepsilon}\rho_a^\sigma(\varepsilon)f(\varepsilon),\cr
B_a^\sigma(\tau)&=&\int^\infty_{-\infty}d\epsilon
e^{-{\rm i}\tau\varepsilon}\rho_a^\sigma(\varepsilon)f(-\varepsilon).
\end{eqnarray}

In general, the current operator is expressed as 
\begin{eqnarray}
j=e\sum_{\sigma,{\bf k}}\sum_{mm^\prime}
v^{mm^\prime}_{\bf k}c^\dagger_{m{\bf k}}
c_{m^\prime{\bf k}},
\end{eqnarray}
where $m$ denotes the band index. 
To derive the expression for the optical conductivity, 
we assume the intra- and interband contribution to be equal
($v^{mm^\prime}_{\bf k}=v_{\bf k}$).
Using the linear response theory, we assume that the momentum 
conservation is violated in real systems by defects and phonon-assisted transitions 
and replace  
$v_{\bf k}v_{\bf k^\prime}\delta_{\bf kk^\prime}$ with 
$v^2/N$ in the current-current correlation function. 
Then, we finally obtain a joint-DOS-like form for the optical conductivity: 
\begin{eqnarray}\label{eq:j-dos}
\sigma(\omega,T)&=&\frac{\pi(ev)^2}{\hbar}\sum_\sigma
\int^\infty_{-\infty}d\nu 
\frac{f(\nu)-f(\nu+\omega)}{\omega}\cr
&& \hspace{-1cm}\times
[\rho^\sigma_{1}(\nu)+\rho^\sigma_{2}(\nu)]
[\rho^\sigma_{1}(\nu+\omega)+\rho^\sigma_{2}(\nu+\omega)
]. 
\end{eqnarray}
In the numerical calculation we set $(ev)^2/\hbar=1$ for simplicity. 

\begin{figure}\vspace{0.5cm} 
\epsfxsize=8cm
\centerline{\epsfbox{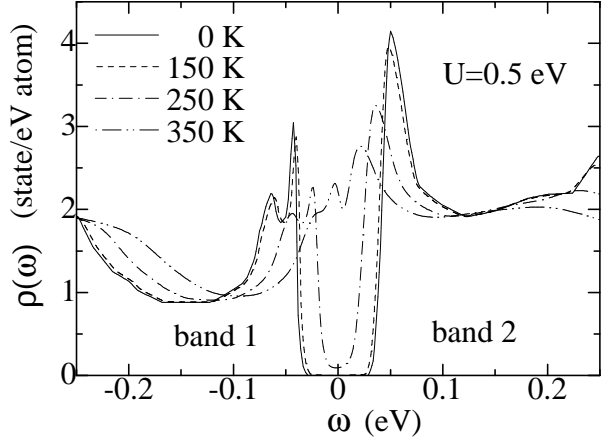}}
\caption{The initial DOS obtained from the band calculation(Ref. 26) at $T=0$. 
At finite $T$, the DOS is strongly temperature dependent 
due to the correlation effects.}
\label{fig:1}
\end{figure}
Figure 1 shows the temperature dependence 
of the quasiparticle DOS for $U=0.5$ eV and $J=0.35U$. 
These values are chosen so as to reproduce the shape and the 
temperature dependence of the optical conductivity spectrum. 
At 0 K, the band 1 is filled and the band 2 is empty, 
so that the correlation effect is absent except the Hartree-Fock contribution. 
At finite $T$, however, the correlation is introduced through the thermally excited 
electrons and holes. 
As a result, the DOS becomes strongly temperature dependent and 
is smeared especially near the gap. 
The gap at 0 K is almost filled up 
at the temperature of the order of its size and 
it is also found that 
the peaks of the gap edges move toward the center of the gap. 

\begin{figure}
\epsfxsize=8cm
\centerline{\epsfbox{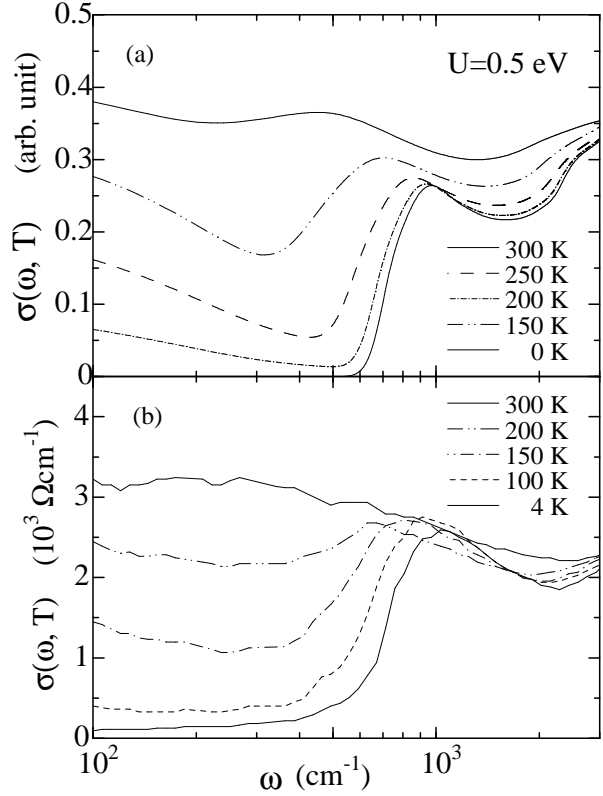}}
\caption{(a)The temperature dependence of the optical conductivity 
calculated with the eq. (\ref{eq:j-dos}). 
(b)The experimental data from Ref. 11. 
The peaks due to phonons observed in the gap are omitted. }
\label{fig:2}
\end{figure}
In Fig. 2(a), the temperature variation of the optical conductivity 
calculated from eq. (\ref{eq:j-dos}) is shown. 
Since the intraband (Drude) contribution does not exist at 0 K, 
only the interband contribution survives and seems to reproduce 
the shape of the spectrum of the experiment at 4 K. 
The slightly rounded gap edge in the calculation is due to the tails at 
the gap edge in the DOS of the band calculation.\cite{Yamada99} 
If we extrapolate the steepest parts of the edge, we obtain the gap 
size as about 75 meV, which is consistent with the gap size of the DOS. 
Since these features reflect the shape of the DOS directly (see eq.(\ref{eq:j-dos})), 
the band calculation using LMTO-ASA by Yamada {\it et al.}\cite{Yamada99} seems to 
give good results about the whole structure of the DOS at $T=0$ 
except for a slightly smaller gap size. 

At finite temperatures, 
the experiment done by Damascelli {\it et al.}\cite{Damascelli97} 
(Fig. 2(b)) shows 
not only that the gap is almost filled up at 300 K, 
but that the increase of the weight in the gap from 100 to 300 K seems to be 
more rapid than that from 4 to 100 K. 
One can also recognize that the peak at the gap edge shifts to 
the lower frequency region as the temperature rises. 
In our calculation (Fig. 2(a)), 
the gap is almost filled up at 300 K as well as 
the rapid increase from 150 to 300 K, 
but a little slower compared to the experiment, which 
may be mainly because the absence of the correlation effect at 0 K 
in the present model. 
Reflecting the correlation effects the peak at the gap edge shifts to lower frequency 
region, as is seen in the experiment. 
In our calculation, however, 
there are dips between the Drude and the interband contributions in contrast
to the experiment. 
This may be caused by the assumption in the previous formulation, where 
we neglected the ${\bf k}$-dependence of the current operator and 
took the intraband (Drude) and interband contribution to be equal
in deriving eq. (\ref{eq:j-dos}).

\begin{figure}\vspace{0.5cm} 
\epsfxsize=8cm 
\centerline{\epsfbox{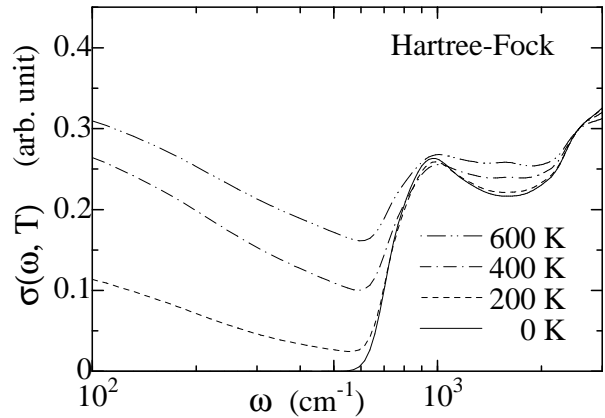}}
\caption{Calculated optical conductivity within the Hartree-Fock approximation.}
\label{fig:3}
\end{figure}
For comparison, the optical conductivity obtained 
from the Hartree-Fock approximation or a rigid band model 
is shown in Fig. 3. 
Since the gap is filled only with the Drude contribution, 
the temperature variation is monotonous and the gap is not efficiently filled 
at a temperature of the order of the gap size as pointed out by 
Fu {\it et al.}\cite{Fu94} 
The peak at the gap edge does not shift. 
These results are consistent with the previous studies,\cite{Fu94,Ohta94} 
but are in disagreement with the experiments. 

In summary, we have calculated the correlation effects using 
the two-band Hubbard model and the DOS obtained 
from the band calculation, and compared the calculated optical conductivity 
with the experiment. 
Although we used an essential but simplified model 
and applied the local approximations in deriving eq. (\ref{eq:j-dos}), 
the calculated optical conductivity reproduced the experiment rather well. 
Thus we determined the gap size (indirect one) as 75 meV when the tail is neglected 
(60 meV within the tail). 
Furthermore, we confirmed that the correlation effect is essential 
to explain the rapid temperature dependence of the optical conductivity, 
and gave a direct explanation for the shift of the peak at the gap edge. 
The latter is another possible sign of the correlation effects. 
Similar effect is also seen in the optical measurement of 
YbB$_{12}$,\cite{Okamura98} 
but the shift is observed on the IR peak instead of 
the peak at the minimum gap edge. 

In the present model the higher order terms of 
the electron correlation effect are quenched at $T=0$. 
The renomalization effects at $T=0$, however, may be also 
important as in the model by Fu and Doniach. \cite{Fu95} 
The optical conductivity of this model is strongly temperature dependent 
as is seen in our previous calculation. \cite{Urasaki98} 
It is also important to take the spin fluctuations\cite{Takahashi79,Saso99} into account 
at higher temperatures or for magnetic properties. 
Furthermore, 
to incorporate the details of the band dispersions 
into the methods of the strongly correlated electron systems 
is our present interest. 

\section*{Acknowledgements}
The authors would like to thank Professor H. Yamada for providing them 
the details of the band calculation (LMTO-ASA) for FeSi and 
for his useful comments.
This work is supported by Grant-in-Aid for Scientific Research No.11640367
from the Ministry of Education, Science, Sports and Culture.

\newpage
\baselineskip=6ex
\noindent
{\bf Figure Captions}

\noindent
Figure 1: The initial DOS obtained from the band calculation(Ref. 26) at $T=0$. 
At finite $T$, the DOS is strongly temperature dependent 
due to the correlation effects.

\noindent
Figure 2: (a)The temperature dependence of the optical conductivity 
calculated with the eq. (\ref{eq:j-dos}). 
(b)The experimental data from Ref. 11. 
The peaks due to phonons observed in the gap are omitted. 

\noindent
Figure 3: Calculated optical conductivity within the Hartree-Fock approximation.
\end{document}